\documentclass[sigconf]{acmart}
\copyrightyear{2024}
\acmYear{2024}
\setcopyright{rightsretained}
\acmConference[ESEM '24]{Proceedings of the 18th ACM / IEEE International Symposium on Empirical Software Engineering and Measurement}{October 24--25, 2024}{Barcelona, Spain}
\acmBooktitle{Proceedings of the 18th ACM / IEEE International Symposium on Empirical Software Engineering and Measurement (ESEM '24), October 24--25, 2024, Barcelona, Spain}
\acmDOI{10.1145/3674805.3690756}
\acmISBN{979-8-4007-1047-6/24/10}

\begin{document}
\title{Rusty Linux: Advances in Rust for Linux Kernel Development}

\author{Shane K. Panter}
    \affiliation{%
    \institution{Department of Computer Science}
    \institution{Boise State University}
    \city{Boise}
    \state{Idaho}
    \country{USA}}
    \email{shanepanter@boisestate.edu}

\author{Nasir U. Eisty}
    \affiliation{%
    \institution{Department of Computer Science}
    \institution{Boise State University}
    \city{Boise}
    \state{Idaho}
    \country{USA}}
    \email{nasireisty@boisestate.edu}

\begin{abstract}

  \textit{\textbf{Context:}} The integration of Rust into kernel development is a transformative endeavor aimed at enhancing system security and reliability by leveraging Rust's strong memory safety guarantees.
  \textit{\textbf{Objective:}} We aim to find the current advances in using Rust in Kernel development to reduce the number of memory safety vulnerabilities in one of the most critical pieces of software that underpins all modern applications.
  \textit{\textbf{Method:}} By analyzing a broad spectrum of studies, we identify the advantages Rust offers, highlight the challenges faced, and emphasise the need for community consensus on Rust's adoption.
  \textit{\textbf{Results:}} Our findings suggest that while the initial implementations of Rust in the kernel show promising results in terms of safety and stability, significant challenges remain. These challenges include achieving seamless interoperability with existing kernel components, maintaining performance, and ensuring adequate support and tooling for developers.
  \textit{\textbf{Conclusions:}} This study  underscores the need for
  continued research and practical implementation efforts to fully realize the benefits of Rust. By addressing these challenges, the integration of Rust could mark a significant step forward in the evolution of operating system development towards safer and more reliable systems.

\end{abstract}




\received[accepted]{22 July 2024}

\maketitle

\section{Introduction}

The 1995 movie Hackers prescient predictions regarding the ease
of breaking into computing systems in cyberspace have come to fruition. The White House Office of
the National Cyber Director (ONCD) released a report calling for the technical community to
proactively reduce the attack surface in cyberspace with a two-pronged approach~\cite{United_States_Gov2024-pp}. First, we need to
address the root cause of many of the most heinous cyber attacks: memory-unsafe programming
languages. Second, we need to establish better cybersecurity quality
metrics so we can have a better understanding of the cybersecurity landscape.

In the ever-evolving landscape of software development, the reliability and security of computer
systems stand as a paramount concern for all parties involved. Modern software is constructed by
building ever more complex abstractions, one on top of the other. Thus, if we aim to have a secure
system, we must start to peel back all the layers and tackle one of the fundamental abstractions in
computer science: the programming language. Programming languages that provide and enforce memory
safety eliminate whole classes of bugs, such as buffer overflows, dangling pointers, and memory leaks
which have been implicated in a myriad of security vulnerabilities and system crashes.

Most operating system kernels are predominantly written in the C programming language with bits of
assembly. C has been favored for its low-level capabilities and performance efficiency, crucial for
kernel development. However, the inherent lack of memory safety in C has led to numerous security
vulnerabilities, including buffer overflows and use-after-free errors, which have plagued operation
system development for decades~\cite{MSRC_Team2019-bf}. Addressing these vulnerabilities is
paramount to enhancing the security and reliability of all operating systems.

Rust, a systems programming language developed by Mozilla, has garnered significant attention for
its strong emphasis on memory safety without sacrificing performance. Rust's ownership model and its
compile-time checks effectively prevent common programming errors that lead to security
vulnerabilities~\cite{Klabnik2024-id}. What makes Rust unique is the fact that it accomplishes these tasks without the use of a garbage collector, making it an ideal candidate for kernel development where safety and performance are critical.

The potential integration of Rust into kernel development represents a significant shift in the
landscape of operating system development. This shift prompts a comprehensive evaluation of both the
opportunities and challenges associated with using Rust as such a foundational component of modern
computing. Therefore, we aim to synthesize existing research and practical
experiences related to the use of Rust in the kernel, providing a detailed understanding of
the current state of this emerging field.

This reflection paper presents a systematic literature review (SLR) focusing on strategies and methodologies
for integrating Rust into one of the most fundamental areas that are typically dominated by unsafe
languages, the operating system kernel. We aim to provide a comprehensive overview of
existing research, identify gaps, and suggest future directions in this domain. Through a rigorous
search process, we synthesized relevant studies and extracted key findings to offer insights into
effective approaches for ensuring memory safety when working closely with hardware.

 \begin{figure*}
    \centering
     \includegraphics[width=7in]{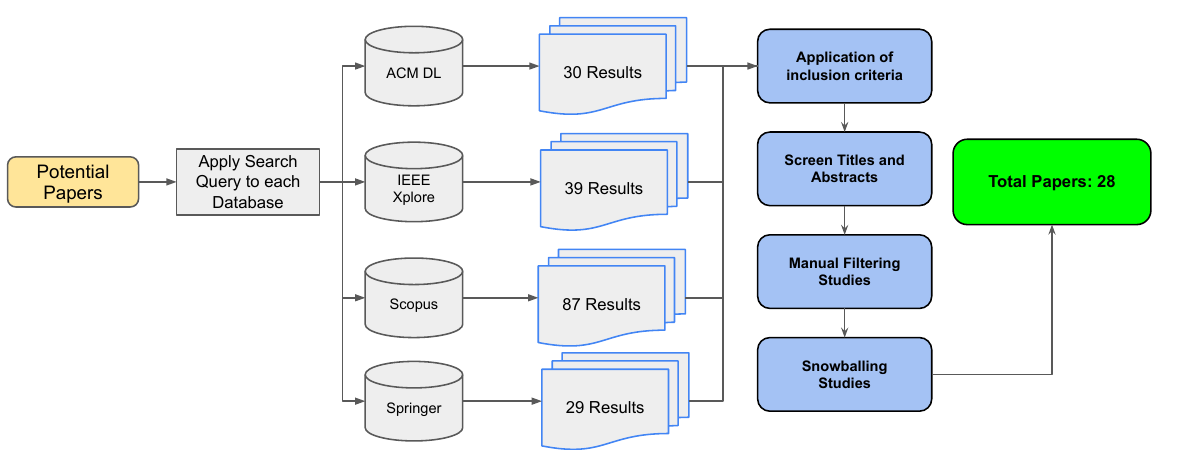}
     \caption{Process Diagram for Paper selection}
     \Description{The process that we followed for selecting papers}
     \label{fig:process}
 \end{figure*}

\section{Research Methodology}

For our research methodology, we followed the Kitchenham and Charters~\cite{Stuart2007-cc} methodology to conduct a SLR in software engineering. We divided our review into three discrete phases, planning the review, conducting the review, and reporting the review results. The following sections detail our review process and are diagrammed in figure~\ref{fig:process}.

\subsection{Planning}

To begin our study, we must first confirm the need for a SLR. The recent report released by
ONCD~\cite{United_States_Gov2024-pp} has conveniently done this job for us by compiling a report
detailing the need for research in the domain of memory safety. While the ONCD report detailed a two-pronged approach, for this reflection paper, we will be focusing on a memory-safe programming language,
specifically Rust. While there are many modern memory-safe programming languages available for
software developers to use, Rust is one of the few languages that is feasible to use when developing
operating system kernels due to its lack of runtime and garbage collector~\cite{Ojeda2021-ul}. The Linux kernel is currently
in the early stages of adding real support for
Rust~\cite{The_kernel_development_community_undated-iw} thus, we will focus on Rust as a primary
candidate to replace the aging C programming language.

\subsection{Research Questions}
\label{sec:researchQuestions}

We have defined the following research questions:

\begin{itemize}
    \item \textbf{RQ1: What are the existing approaches for implementing operating system kernels in Rust?}

    Although Rust in the Linux kernel is recent, we want to see what approaches researchers have explored in any operating system. Experiential and research kernels can provide new ideas that can be integrated into the Linux kernel.
    \item \textbf{RQ2: What are the performance implications of using Rust for operating system kernel development?}

    The C programming language has the advantage of 40 years of improvements and lessons learned regarding how to write fast code. Rust, being a new language, still has many unexplored areas regarding how to write fast code, especially in kernel space. Therefore, we want to explore the performance implications of Rust for kernel development in terms of throughput, latency, and resource utilization.
    \item \textbf{RQ3: What are the major challenges and limitations when developing operating system kernels in Rust?}

      Kernel space presents unique challenges for the software engineer using Rust. Rust was not designed to implement kernels and is more restrictive than C. Thus, we want to identify if there are any stumbling blocks that have already been found when using Rust in kernel development.
    \item \textbf{RQ4: What are the lessons learned when developing operating systems kernels in Rust?}

    Besides technical challenges and limitations, developers' perceptions are very important in software development. Therefore, through this research question, we want to find developers' reflections on what went well and what didn't go well while developing kernels in Rust.
\end{itemize}

\subsection{Data Collection}

\subsubsection{\textbf{Search Strategy.}}

We employed a robust multi-step search strategy across four databases: ACM Digital Library,
IEEE Xplore, Scopus, and Springer Link to find all the current research.
We leveraged the advanced search features of all three databases to search the titles and abstracts for
keywords using boolean search operators.

\subsubsection{\textbf{Search criteria.}}

We searched all four databases with the keywords listed in table~\ref{tab:keywords} between January
1, 2019 and April 1, 2024. We choose to look at only the previous 5 years of research in order to capture the
bleeding-edge research that is currently being done. While both ACM and IEEE had mutually exclusive results, Scopus had 44
duplicates that needed to be removed and Springer Link didn't have any new papers that were not also in ACM or IEEE. We then used Google Scholar to do forward and backward snowballing.

\begin{table}
\begin{tabular}{|| p{2cm}| p{4cm} | p{1cm} ||}
 \hline
 Database & Query & Results \\
 \hline\hline
 ACM  & (Abstract:("operating system" OR Kernel OR linux OR OS) AND Abstract:(Rust)) OR (Title:("operating system" OR OS OR kernel OR linux) AND Title:(Rust))  & 30 \\
 IEEE & ("All Metadata":"operating system" AND "All Metadata":rust) OR ("All Metadata":kernel AND
 "All Metadata":rust) OR ("All Metadata":linux AND "All Metadata":rust) & 39 \\
 Scopus & TITLE-ABS-KEY ( ( "operating system" OR kernel OR linux ) AND rust ) AND PUBYEAR > 2018
 AND PUBYEAR < 2025 AND ( LIMIT-TO ( SUBJAREA , "COMP" ) ) AND ( LIMIT-TO ( LANGUAGE , "English" ) )
 & 87 \\
 Springer Link & Operating System Rust Kernel (Conference Paper, Article, Research article) Subdiscipline: Software engineering/programming and operating systems & 29 \\
 \hline
\end{tabular}
\caption{Search Queries used for each Database.}
\label{tab:keywords}
\end{table}

\subsubsection{\textbf{Inclusion and exclusion criteria.}}

\begin{table*}
    \caption{Approaches and Methodologies for Rust in the Kernel}
    \begin{tabular}{||l|l|l||}
    \hline
    Approach & Papers & Operating System in Rust\\
    \hline\hline
    Monolithic  & \cite{The_kernel_development_community_undated-iw}, \cite{Li2019-ru}, \cite{Miller2021-pg}, \cite{Oikawa2023-ms} & \href{https://docs.kernel.org/rust/}{Linux kernel v6.1+}\\
    Micro-kernel & \cite{Chen2023-wb},\cite{Liang2021-bo}, \cite{Liu2024-xe}, \cite{Narayanan2020-gs}, \cite{Narayanan2019-fd} & Atmosphere, Redox, Redleaf\\
    Embedded & \cite{Culic2022-bk}, \cite{Vishnunaryan2022-yd} & \href{https://github.com/tock/tock}{Tock}, \href{https://hubris.oxide.computer/}{Hubris}, \href{https://www.drone-os.com/}{Drone}, \href{https://bern-rtos.org/}{Bern}, HarSaRK \\
    Unikernel & \cite{Lankes2019-cm},  \cite{Boos2020-zh}, \cite{Ijaz2023-da}, \cite{Sung2020-bb}  & RustyHermit, Theseus \\
    Exokernel & \cite{Li2024-yb} & W-Kernel \\
    \hline
  \end{tabular}
  \label{tab:RQ1}
\end{table*}

Following the guidelines outlined by Kitchenham and Charters~\cite{Stuart2007-cc} we set the
inclusion and exclusion criteria based on our research questions outlined in
section~\ref{sec:researchQuestions}. We only considered papers that are written in English and published
in conferences, journals, and workshops. The published papers should
describe using the Rust programming language for either developing a new kernel, extending an
existing kernel, or authoring drivers. We included any type of kernel architecture, including
monolithic kernel, microkernel, or unikernel in both the embedded and non-embedded space, as long as
the paper is using the Rust programming language in some way. Papers that describe solutions that
reside 100\% in user space or papers that relied too heavily on a component written in C were excluded. Early efforts to use Rust in kernel development
were impacted by changes and updates in both the Rust language and compiler and relied on unstable
features, which were thus excluded from this review.

\subsection{Data Analysis}

Once we had our initial collection of papers from the databases, we merged all duplicate papers into one record and then started our first pass, which consisted of reading the title and abstract and either marking the paper as \textit{include} or \textit{exclude}. Once we completed the initial screening, we were left with 22 papers that needed an in-depth reading and analysis. We found an additional 6 papers during forward and backward snowballing that were also reviewed, bringing the total number of papers to 28. The first author manually reviewed all 28 papers, built the first dataset, and conducted the analysis. The second author then verified all the data and results. Since the results were straightforward and came from a comparatively small number of papers, there were no discrepancies or missing results to solve.

\section{Results}

In this section, we present our findings. We address each research question from \textbf{RQ1} - \textbf{RQ4}.

\subsection{RQ1: Existing Approaches}

The integration of Rust into operating system kernel development has garnered significant attention
due to Rust's promise of memory safety and concurrency guarantees. Several approaches and
methodologies have emerged to leverage Rust for kernel development, each addressing different
aspects of kernel functionality and integration.  We found several approaches that researchers are
taking to integrate Rust into kernel development. The first approach is a greenfield approach
where the operating system architecture is redesigned and built from the ground up in order to fully
leverage the Rust programming language~\cite{Culic2022-bk, Boos2020-zh}. The second approach is the incremental integration of Rust
into existing kernels, such as the Linux
kernel~\cite{The_kernel_development_community_undated-iw, Miller2021-pg, Oikawa2023-ms}. This methodology involves gradually
rewriting or augmenting specific components of the kernel with Rust while maintaining compatibility
with the existing C codebase. Finally, researches are porting existing C kernels over to Rust and then evaluating the differences. Table~\ref{tab:RQ1} summarizes our findings regarding what existing
approaches researchers are using to integrate rust into the kernel.

For the first approach, Boos et al.~\cite{Boos2020-zh} created an experimental operating system
named Theseus that operates in a single address space and single privilege level and uses properties
of the Rust programming language to realize isolation instead of relying on hardware. This novel
approach pushes some of the roles that a traditional operating system would take onto the
compiler. The design of Theseus uses a novel cell-based structure where ownership of memory and
resources is enforced by the compiler, thus avoiding these checks during runtime to get improved
performance. In addition to memory safety, Theseus has a goal of addressing the state spill
problem. State spill happens when a single service in an operating system can harbor a state change
induced by interacting with other services. That state change can eventually cause a system crash or
system instability at some point in the future, making it very difficult to track down the problem
due to the separation between when the problem occurred and when it was observed. Theseus OS is an
operating system that was designed to match the Rust language instead of the more traditional route
of matching the hardware. Li and Sato~\cite{Li2024-yb} explored using Rust to write an exokernel named W-Kernel. The authors proposed a novel architecture for an OS by embedding a WebAssembly (WASM) runtime into the kernel that can run programs written in any language that can compile down to WASM.

For the second approach, researchers take a more traditional micro-kernel implementation~\cite{Chen2023-wb, Liang2021-bo}. By pushing most kernel functionality
to user space, the authors limited the surface area that needs to be proved. Chen X et al.~\cite{Chen2023-wb} took
their implementation one step further by formally verifying their new micro-kernel named Atmosphere
by using both the liner type system of Rust in combination with an SMT solver. They were
able to get a 7.5:1 proof-to-code ratio, which is higher than other formerly verified micro-kernels
SeL4 and CeriKOS both of which have proof code ratios of 19:1 and 20:1, respectively. While this
approach was novel, drivers running in userspace are still not formally verified and don't have to be
written in Rust.

Finally, researchers have taken existing unikernels written in C and ported them to Rust. Unikernels are designed to do one thing
and one thing only, thus limiting the complexity that a general-purpose operating system has to
deal with~\cite{Madhavapeddy2014-zw}. This reduced complexity allowed  Lankes et al.~\cite{Lankes2019-cm} to take HermitCore and rewrite it in Rust. The new
Unikernel named RustyHermit consists of only 3.27\% unsafe Rust, with the rest of the code base consisting of safe Rust, dramatically decreasing the surface area where memory errors could originate.

\subsection{RQ2: Performance Implications}

\begin{table}
    \caption{Performance Implications of Rust in the Kernel}
    \begin{tabular}{||l|l|l||}
    \hline
    No. & Implication & Studies that Reported the challenge\\
    \hline\hline
    1 & Performance & \cite{Gonzalez2023-ek}, \cite{Li2024-be}, \cite{Ma2023-ef}\\
    2 & Throughput & \cite{Gonzalez2023-ek}\\
    3 & Latency & \cite{Culic2022-bk} \\
    \hline
  \end{tabular}
  \label{tab:RQ2}
\end{table}

Comparing performance, throughput and latency between different systems with different architectures is very difficult. For
example, several of the operating systems written in Rust are a complete rethinking of
how an operating system is designed~\cite{Boos2020-zh} while others run entirely in kernel space~\cite{Lankes2019-cm}. Additionally, direct comparisons between a Monolithic kernel, Micro-kernel, and Unikernel, or
comparisons between a Real Time Operation System (RTOS) and a General Purpose Operating System, are
not directly meaningful due to the vastly different goals of each system and the overhead that is imposed by the hardware~\cite{Arnott2012-fl}. Therefore, we will focus more on the overall efforts that are specific to the Rust programming language and the challenges
presented. We summarize our findings in table~\ref{tab:RQ2}.

Culic et al.~\cite{Culic2022-bk} looked at latency issues in Tock. Tock is a new operating
system written in Rust that is designed to run on embedded systems but does not provide real-time capabilities. The
authors attempted to add real-time capabilities by integrating eBPF into the Tock kernel to improve
the interrupt handler's response time. The authors found that early work (still in the prototype
stage) lowers the response times of the system and interrupts response times 3x.

Gonzalez et al.~\cite{Gonzalez2023-ek} explored using the Rust for Linux Project to implement a
native UDP driver in Rust in order to explore the performance of the Rust programming language. The
authors were able to get a basic driver working with performance only slightly slower than C using
the Rust for Linux (RFL) project. The RFL project is still too immature to get a full driver up and
running, but it is at a stage where researchers can start experimenting with different approaches.

Li et al.~\cite{Li2024-be} explored the feasibility of using Rust in kernel space. The authors took
an existing component, the Out of Memory (OOM), and implemented a replacement using the Rust
programming language. The non-encapsulated interface Rust component which was almost identical to
the original C component, only introduced a 0.7\% overhead. The encapsulated Rust component, on the
other hand, added a 3\% performance overhead.

\subsection{RQ3: Challenges and Limitations}

\begin{table*}
    \caption{Challenges Unique to the Rust Programming Language}
    \begin{tabular}{p{1cm} p{5cm} p{10cm}}
        \hline
        No. & Challenge & Description\\
        \hline
        1 & Binary Size~\cite{Ayers2022-sf}  &
        \begin{itemize}
            \item Deeply ingrained monomorphization which increased the size of the rust binaries
            \item Compiler optimizations are not as mature as some C based compilers, thus increasing binary size
            \item Hidden data structures and data
            \item Sub-optimal compiler generated support code
        \end{itemize}
        \\
        \hline
        2 & Missing Features~\cite{Burtsev2021-mh} &
        \begin{itemize}
            \item Supporting trait bounds on functions and closures with any number of arguments
            \item Expose type information in procedural macros
            \item Support a collision-free, unique type identifier

            \item Support typed assembly language for Rust
            \item Support trusted build environments
            \item Provide software-only stack guard with extensible probing interface
            \item Develop zero-copy serialization of “plain-old” data structures
        \end{itemize}
        \\
        \hline
        3 & Soundness~\cite{Klimt2023-ob} &
        \begin{itemize}
            \item Unsynchronized Global State - any use of mutable statics is unsafe
            \item C-Style Abstractions - Use rust style abstractions to properly encapsulate internal unsafe usage of raw pointers
            \item Aliasing Mutable References - Giving out raw pointers to memory that is also referenced mutably
            \item Re-implementing Memory Access - instead of accessing specific memory regions in assembly, creating and using references to the whole region is preferable
        \end{itemize}
        \\
        \hline
        4 & Panics \cite{Ma2023-ef},\cite{Burtsev2021-mh} &
        \begin{itemize}
            \item Support extendable, no\_std unwind library
            \item Stack unwinding in embedded environments
        \end{itemize}
        \\
        \hline
        5 & C Interop \cite{Miller2019-xm}, \cite{Li2021-xo} &
        \begin{itemize}
            \item Kernel interfaces, while designed for extensibility, are not designed for type safety
            \item Hybrid Code Flow. The Rust compiler can not track ownership when switching between modules written in C and Rust
        \end{itemize}
        \\
    \hline
  \end{tabular}
  \label{tab:RQ3}
\end{table*}

The development of operating system kernels in Rust introduces several unique challenges and
limitations compared to more traditional languages like C. These challenges arise from Rust’s strict
safety guarantees, its relatively recent adoption in systems programming, and the inherent
complexities of kernel development~\cite{Ojeda2021-ul}. This section discusses these challenges, limitations, and the
lessons learned from various projects and research efforts. We summarize our findings in
table~\ref{tab:RQ3}.

A primary concern for programming languages that are used for operating system development is the
size of the binary especially in an embedded environment, we must
be careful to keep the size of the Rust component as close as we can to the C component so we can
still run on the same hardware. Li et al.~\cite{Li2024-be} found that there was only a 0.06\% size increase when compared to the original C implementation.  Ayers et al.~\cite{Ayers2022-sf} focused
on reducing the size of binaries produced by the rust compiler while working with Tock OS. They were
able to identify several  causes of binary growth that are specific to the Rust programming language
and have the following 5 recommendations when using Rust in a size-constrained environment:
\begin{itemize}
  \item Minimize Length + Instantiations of Generic Code
  \item Use Trait Objects Sparingly
  \item Don't Panic
  \item Carefully Use Compiler Generated Support Code
  \item Don't use static mut
\end{itemize}

Burtsev et al.~\cite{Burtsev2021-mh} explored what is missing in the Rust programming language to
help solve the isolation problem in an operating system. Currently, Rust lacks the ability to express
isolation in the heap without external support. For example, the RedLeaf experiential operating
system, which was written in Rust, relied on a complex interface definition language (IDL) to enforce
isolated heaps. The paper enumerates several properties of the Rust Language that could help with
isolation with regards to operating system development that are detailed in table~\ref{tab:RQ3}. The
authors argue that with these changes or inclusions to the Rust programming language developing
operating system kernels would be much easier and safer.

\subsection{RQ4: Lessons Learned}

Klimt et al.~\cite{Klimt2023-ob} details the lessons learned and challenges when implementing
Theseus. First, they found that it is impossible to write a complete operating system in 100\% safe
Rust. For example, when writing a memory management system, raw pointers must be used to
modify the hardware. Despite the limitations imposed by the hardware, the authors describe how
Theseus leveraged intralingual design to maximize the compiler's role in enforcing correctness. By
leveraging Rust's type system and borrow checker memory safety, correct ownership transfer can be
achieved at a higher level than what could be done in C. The authors also detailed some of the
limitations of intralingual design, such as not being as expressive as many other formal verification
techniques due to the limited invariance that the type system can enforce. One of the most
important lessons learned was the insight that a linear type system itself cannot guarantee
the uniqueness of the resource represented, such as when a memory resource may overlap. The authors
introduce the idea of using a hybrid approach of verification where they leverage both the linear
type system and an SMT solver. The authors also explore the bootstrapping problem with rust
systems. In operating systems written in C, the kernel provides the ownership root to applications running on
top. Finding a new ownership root is an open research question with regard to using Rust for
operating system development.

\section{Threats to Validity}

\textit{\textbf{Internal Threats.}} Potential internal threats include:

\begin{itemize}
  \item \textit{Quality of Studies:} The methodological rigor of the included studies varies, and some may
    suffer from design flaws or biases that were not adequately controlled for, thus affecting the
    reliability of their findings.
  \item  \textit{Selection Criteria:} The inclusion and exclusion criteria for selecting relevant studies might inadvertently bias the review towards certain types of research, such as those reporting successful integration while underrepresenting studies detailing challenges and failures.
\end{itemize}

\textit{\textbf{External Threats.}} Threats to external validity include:

\begin{itemize}
  \item \textit{Context-Specific Findings:} Many studies focus on specific kernel modules or use-cases, which
    may not be representative of the broader kernel environment. The success of Rust in isolated
    components does not necessarily translate to the entire kernel.
  \item \textit{Temporal Changes:} The rapidly evolving nature of both the Rust programming language and the
    Linux kernel means that findings from older studies may no longer be applicable, as improvements
    and changes in both domains can alter the landscape significantly.
    \item \textit{Definition Ambiguity:} The term ``Kernel development in Rust'' encompasses a wide range of activities,
    from minor module development to complete subsystem rewrites. Variations in how researchers and
    developers interpret this integration could lead to inconsistent findings.
\end{itemize}

\textbf{\textit{Conclusion Threats}.} Potential conclusion threats include:

\begin{itemize}
  \item \textit{Heterogeneity of Studies:} The diverse methodologies, metrics, and contexts of the included studies
    can lead to challenges in synthesizing findings and drawing unified conclusions. For example, it
    is very difficult to directly compare a micro-kernel, Unikernel, and hybrid Kernel.
  \item \textit{Reviewer Bias:} Personal biases of the reviewers in interpreting data and making judgments
    about study quality and relevance may skew the results
\end{itemize}

We made efforts to minimize this through predefined criteria and multiple reviewers, but some
degree of subjectivity is inevitable.  In addressing these threats, we employed rigorous methods for
study selection, data extraction, and analysis, and we remain transparent about the
limitations. Despite these threats, our results provide a valuable synthesis of the current
state of research on the use of Rust in kernel development, highlighting both its potential benefits
and the challenges that need to be addressed.

\section{Discussion \& Conclusion}

The integration of Rust into  kernel space represents a significant evolution in operating system
development, promising to address long-standing issues related to memory safety and system
reliability. We synthesized the current state of research and
practice concerning the use of Rust in multiple types of kernels, including the venerable Linux
kernel, highlighting both the progress made and the challenges that remain.

Our review indicates that Rust's strong guarantees of memory safety, enabled by its ownership model
and strict compile-time checks, offer a compelling advantage over traditional C-based kernel
development. These features have the potential to reduce common vulnerabilities such as buffer
overflows and use-after-free errors, which are prevalent in C and have historically led to critical
security exploits. Several case studies and prototype implementations have demonstrated that Rust
can be successfully integrated into the kernel, providing safer interfaces and reducing the
incidence of memory-related bugs without incurring significant performance penalties.

However, the adoption of Rust for kernel development is not without its challenges. Truly, the
biggest challenge is existing codebases and mountains of legacy code. With extensive use of C and
reliance on specific C idioms and low-level programming techniques, researchers have substantial
integration hurdles to overcome. Any efforts to rewrite substantial portions of an existing kernel
such as Linux in Rust, are constrained by the need for interoperability with existing C code and the
necessity to maintain the kernel's performance characteristics. Furthermore, the operating system
development community must reach a consensus on Rust's role and ensure that sufficient tooling,
documentation, and support are available for developers.

In conclusion, while Rust's incorporation into the kernel space is still in its nascent stages, the
initial results are promising. The potential for enhanced security and stability aligns with the
long-term goals of kernel development, and continued research, coupled with practical implementation
efforts, will be crucial in realizing these benefits. Future work should focus on addressing
integration challenges, refining interoperability mechanisms, and expanding the body of empirical
evidence on Rust's impact within the kernel environment. As the community navigates these
challenges, the evolution of Rust in the kernel may well mark a transformative period in the pursuit
of safer, more reliable operating systems.

\bibliographystyle{ACM-Reference-Format}
\bibliography{paper.bib}

\end{document}